# The optical applications of 3D sub-wavelength block-copolymer nanostructured functional materials

Z. L. Poole,[a] A. Yan,[a] P. R. Ohodnicki,[b] K. P. Chen[a]

A method to engineer the refractive indices of functional materials ($TiO_2$, $ZnO$, $SnO_2$, $SiO_2$), by nanostructuring in the deep sub-wavelength regime (<20nm), is presented. Block-copolymer templating combined with a wet processing route is used to realize 3D functional nanostructures with continuously adjustable refractive indices from 1.17 to 2.2. Wet processing accessed refractive index engineering can be applied to address a variety of realizability concerns in attaining design specified refractive index values and refractive index gradients in 1D, 2D, and 3D that arise as the results of optical design techniques such as thin film optimization methods, transformation optics and conformal mapping. Refractive index optimized multi-layer anti-reflection coatings on crystalline silicon, which reduce light reflections from 38% down to ~3% with a wide angular span, are demonstrated with the developed wet processing route. A high temperature oxygen free fiber optic hydrogen sensor realized by accessing nano-engineering enabled refractive indices is also presented. The functionality of the sensor is characterized with a fiber Bragg grating, transmission based interrogation, and optical frequency domain reflectometry. The latter demonstrates the potential of the developed sensor for the detection of chemical gradients for applications such as in high temperature hydrogen driven fuel cells.

## Introduction

Nature provides a limited availability in the refractive indices of functional optical materials. It is often the case that designs have to be formulated around existing material properties which are sparsely distributed in a limited range. In optics and photonics, the refractive indices of materials are specifically chosen to obtain a desired functionality as they define the working relationship between light and matter.

Transformation optics and quasi conformal-mapping are an upcoming and interesting design methodology in optics and photonics[1]. With this technique a variety of designs have been carried out to allow for the unconventional manipulation of light for practical applications, such as in light wave circuits and silicon photonics[2]. It is a tensor based design method[3,4] which, in most cases, provides designs that require strong 3D refractive index gradients. These are difficult if not impossible to realize with conventional techniques at optical frequencies.

From theory, an optimal anti-reflection coating would consist of a material with a refractive index that continuously varies from the substrate to the background (1D refractive index gradient)[5,6]. For single crystal silicon, this would require having materials with refractive indices continuously distributed between 3.5 and 1. The realization of refractive index gradients over large index spans for optical frequencies is out of reach with standard manufacturing techniques. A number of methods have been demonstrated such as nano-arrays[7-9], surface textures[10-12], anodic alumina[13], lithography and wet etching[14], and a combination of oblique angle deposition with sputtering techniques[5,15,16] to address the need of improving anti-reflection technology. Many of these are based on technologies such as e-beam, CVD, and sputtering that do not scale well and are highly restrictive in deposition. Currently, the most used anti-reflection coatings are still single layer $Si_3N_4$ and $TiO_2$ which reduce reflections to ~18% over the solar conversion window[15].

Optical fiber is a highly sought after sensory platform for numerous application due to its small form and its ability to withstand harsh conditions such as high temperatures and corrosive environments for which electronic components are highly susceptible[17,18]. A number of sensing applications for optical fiber already exist[17,19-21] but, the inert nature of silica limits its detection capability. Incorporating sensory materials with optical fiber (e.g. evanescent wave configuration) is a practical way to extend the sensory potential of this platform for the detection of numerous biological and chemical species. A number of advances have been already explored including thin palladium films for hydrogen sensing[22,23] and films of various oxides and other materials for the detection of a number of chemical species[24-29]. Response types from absorptive based[27], to refractive index based[25], to surface plasmon resonance based[18,30] have been explored.

Over the past two decades, metal oxide type materials have undergone a number of developments for sensory applications. As such, a variety of techniques exist to tailor the sensory capability of functional materials by enhancing the magnitude of the response, the speed of the response, and the type of the response to a great variety of chemical and biological species[31-35]. Noble metal promoters such as Al, Au, Pt, and Pd and nanostructuring with feature dimensions comparable with the sensory modulated depth are known to improve sensory performances both in response magnitude and rate[31,35-37]. Although, some exploration has already shown that metal oxide thin films can be combined with optical fiber for sensory applications, there exists the need to quantify the functionality of this combination and to optimize the compatibility for maximum sensory responses. The typical refractive indices of sensory metal oxides are greater or equal to ~2, whereas the refractive index of silica, typical material of optical fiber, is

~1.46. This large refractive index difference poses difficulties and constraints in combining advanced functional sensory materials with the optical fiber sensory platform.

In this paper a method to engineer the refractive indices of functional materials such as $TiO_2$, $ZnO$, $SnO_2$, and $SiO_2$ in the range of 1.17 to 2.2, using deep sub-wavelength 3D block-copolymer based nanostructuring, is presented. The method employs a scalable low-cost solution processing approach that is compatible with a variety of deposition methods from spin casting, to dip coating, and to spray coating. The developed method is first applied to realize low-cost, and easily scalable refractive index optimized multilayer anti-reflection coatings on crystalline silicon with a demonstrated reduction in the reflected light from ~38% down to ~3%, in a wide angular span. The omnidirectional nature of gradient based anti-reflection coatings could eliminate the need for solar tracking technologies. Functional materials with engineered refractive indices that are below 1.46 can be easily integrated with optical fiber, extending and enhancing its sensory potential for the detection of numerous chemical and biological species. A refractive index engineered palladium-doped $TiO_2$ nanomaterial coated D-shaped evanescent-wave optical fiber sensor, for the high temperature detection of hydrogen in an oxygen free environment, is demonstrated. With a combination of an incorporated high temperature stable fiber Bragg grating and transmission based interrogation, the sensory response of the developed sensor is characterized. Optical frequency domain reflectometry type distributed sensing is performed on the manufactured Pd-doped $TiO_2$ optical fiber based hydrogen sensor to demonstrate the potential for the detection of chemical gradients that may exist in applications such as in solid oxide fuel cells. The developed solution-based low-cost refractive index engineering technique is then projected to have a variety of potential applications in realizing strong refractive index gradients in 1D, 2D and 3D geometries for applications in photonics and transformation optics.

**Tailoring the Refractive Indices**

The idea to manipulate materials to tailor their dielectric permittivity dates back to the Clausius-Mosotti relation(1850)[38, 39] and later to the Maxwell-Garnet (1904)[40] and the Bruggeman (1935)[41] effective medium theories. In the quasi-static regime, where sufficient separation exists between the wavelength of light and the geometric features of the compositions (~λ/10), these concepts can be realized by modulating the type of the material between two or more components in 3D space. In essence, these formulations establish a link between a microscopic quantity (atomic polarizability) and a macroscopic quantity (dielectric constant) when materials are mixed together in the deep sub-wavelength regime(<20nm)[42]. The material components could be modulated in lower dimensions (nanorods, etc.) however, these would possess strong anisotropic properties in the refractive indices.

Despite the relatively simple nature of these averaging schemes, they provide powerful methods to engineer the optical properties of materials at the near atomic scale. When combined with scalable nano-manipulation methods, they can be used to provide practical solutions to improve on a variety of existing methods and to develop new applications.

The requirement of controlled and scalable 3D nanostructuring of functional materials in the 20nm regime, to examine the realizability of the established effective medium theories, gives way to only one of the existing methods, which is templating by block-copolymers. A rich variety of structures can be obtained by templating with block-copolymers with features in the 5-100nm range[43-45]. For exploring the 3D structuring of functional materials in the deep sub-wavelength regime (<20nm) a non-ionic hydrophilic triblock-copolymer, Pluronic F-127, is used. This triblock-copolymer is known to have a higher temperature stability and to form 3D structures with the metal alkoxide wet processing route[45-48]. After deposition and annealing, the block-copolymer metal-alkoxide composite is transformed into a 3D functional material matrix, providing near isotropic refractive indices. Varying the molecular ratio between the block-copolymer and the metal alkoxide provides variations in the size[49] and distribution of the formed morphologies. This gives rise to controllable volume fraction variations on the <20nm scale that directly influence the refractive indices.

**Anti-Reflection Coating Design**

Quarter-wave thin film stacks are very efficient at minimizing reflections but at the high cost of greatly reduced angular and wavelength spans. Optimal anti-reflection coatings predicted by theory that minimize light reflections globally over a large angular and wavelength span require materials with continuous refractive index profiles from the substrate to the background refractive index (1D refractive index gradients). Profiles that minimize reflections have been previously explored analytically[50] and by numerical optimization methods, such as the genetic algorithm[5, 51, 52], the simulated annealing algorithm[53], and the ant colony algorithm[54]. Practical approximations of the continuous profiles are multi-layered coatings designed by the rigorous transfer matrix method[55] coupled with numerical optimization techniques[5, 52].

In this paper, the multi-layer anti-reflections coatings were designed by the rigorous transfer matrix method coupled with the genetic algorithm and the found solutions were verified with the simulated annealing algorithm to ensure consistency. The transfer matrix method was used to determine the amount of the reflected light as a function of wavelength, angles of incidence, polarization state, and the thickness and refractive index of each layer. Bruggeman's effective medium theory was used to link the volume fraction of the nanomaterial air 3D mixture with the refractive indices[42].

$$f\left(\frac{\varepsilon_1 - \varepsilon_{eff}}{\varepsilon_1 + 2\varepsilon_{eff}}\right) + (1-f)\left(\frac{\varepsilon_2 - \varepsilon_{eff}}{\varepsilon_2 + 2\varepsilon_{eff}}\right) = 0$$

During optimization solutions were evaluated with a fitness function providing the average reflectance over wavelength and angles of incidence[52]. As previously suggested, the weight function w can be used to preferentially adjust the parameters to account for, for example, variations in light intensity in the solar spectrum as a function of wavelength and angle of incidence or in the internal/external quantum efficiency of solar cells[52].

$$R_{avg} = \frac{1}{\Delta\lambda\Delta\theta}\int_{\lambda_{min}}^{\lambda_{max}}\int_{\theta_{min}}^{\theta_{max}}\frac{1}{2}w(\lambda,\theta)\cdot(R_{TE}+R_{TM})d\theta d\lambda$$

The nominal refractive indices ($\tilde{n} = n + ik$) for crystalline silicon, titanium dioxide, and silicon dioxide were obtained from the literature[56-58]. Titanium dioxide is used in direct contact with the substrate, due to its high refractive index. Silicon dioxide is used to model successive layers, due to its low refractive index. Simulations were performed over the indicated wavelength and angular ranges with values in parenthesis indicating the simulation results for the reduced range (**Table 1**). Horizon effects eliminate the need to consider angular spans over 75° from the normal. The reduced range simulations were performed as these will be manufactured and characterized, and in this simulation current limits with the developed method of refractive index engineering for a crystalline silicon substrate are included.

**Integrating with Optical Fiber**

It has been shown in prior works that thin films of high refractive index materials can be successfully integrated with optical fiber for sensory applications. The functionality of such combinations is not well quantified since it is known that the placement of a high refractive index material in close proximity of a low refractive index fiber core will severely degrade and potentially destroy the guiding properties of the fiber[59, 60]. Numerical simulations were performed to uncover the working principle of previously demonstrated combinations and to examine the currently proposed integration scheme (**Figure 2**), the engineering of the refractive indices for on-fiber compatibility designs. In a single-mode optical fiber the wave guiding condition requires the propagation constant of the fundamental mode to be greater than that of the fiber cladding and smaller than that of the fiber core ($n_{core}k_o > \beta > n_{cladding}k_o$, $\beta = k_o n_{eff}$)[61]. Replacing part of the cladding material with a sensory film in direct contact with the fiber core will alter the guiding properties of the fiber ($\beta, n_{eff}$)[59]. Assuming that the sensing element is comprised of a finite portion of the waveguide, it is advantageous to maintain a constant effective propagation constant along the entire length of the fiber to maximize overall useful light transmission. As such finite element modal analysis was employed to determine the values of refractive indices and associated film thicknesses that return the propagation constant ($n_{eff}$) of the sensing section to the nominal value. **Figure 3** indicates the film refractive index and thickness combinations that satisfy this condition. The simulation obtained trends are consistent with previous demonstrations. In addition a confinement factor analysis was performed to indicate the interaction potential of the evanescent wave with films of various thicknesses and refractive indices. The confinement factor is the integral of the power density in the film with respect to the entire geometry. The results show that a sensory film of refractive index of 1.9 would require a film thickness of 120nm or less in order to maintain the guiding condition in the fiber core. On the other hand, refractive index engineered functional material films with indices below that of the fiber core (~1.47) can be relatively thick while still maintaining the necessary guiding condition. Without accounting for the effects of microstructure on the films sensory response, a film of 2μm in thickness is thereby expected to provide almost an order of magnitude increase in the interaction potential with the evanescent field.

**Table 1**: Multi-layer simulation results for λ of 400-1100(400-700)nm and θ of 0-75(0-60)°

| Layers | 0 | 1 | 2 | 3 |
|---|---|---|---|---|
| v1 |  | 81(84)%TiO$_2$ | 77(51)%SiO$_2$ | 20(40)%SiO$_2$ |
| v2 |  |  | 100(100)%TiO$_2$ | 99(74)%SiO$_2$ |
| v3 |  |  |  | 100(100)%TiO$_2$ |
| h1 |  | 88.45(64)nm | 126(113)nm | 223(200)nm |
| h2 |  |  | 69(56)nm | 107(120)nm |
| h3 |  |  |  | 68(57)nm |
| Ravg | 35(38)% | 13.60(6.64)% | 6.15(2.8)% | 4.70(3.6)% |

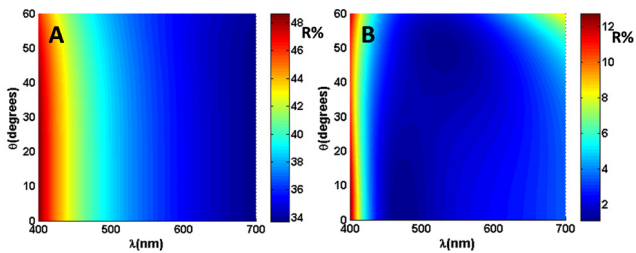

**Figure 1:** Light reflectivity of **A:** bare silicon and **B:** a thickness and refractive index optimized two-layer coating composed of TiO$_2$ and SiO$_2$ nanomaterials as a function of wavelength and angles of incidence.

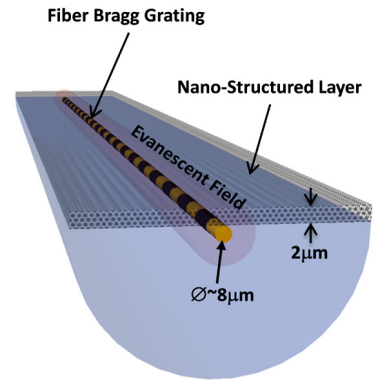

**Figure 2:** Schematic of a functional nanomaterial integrated D-shaped optical fiber with an in-line fiber Bragg grating, in the evanescent wave configuration

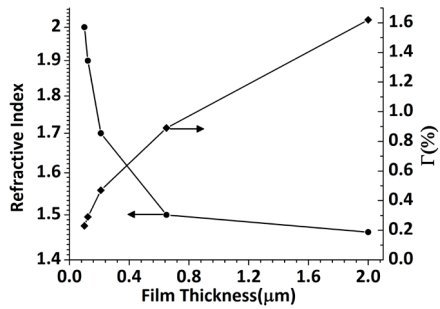

**Figure 3:** Modal analysis scan over refractive indices and the associated film thicknesses that maintain the nominal propagation constant. Right axis showing the associated evanescent confinement factor Γ.

**Expected Sensory Response**

Metal oxide type chemiresitive sensors are well explored and it is known that a conductometric response is to be expected. It is also known that structuring on the scale that is comparable with the sensory modulated depth (~1-10nm, proportional to the Debye length) can provide significant enhancements in the magnitude and rate of the conductometric type responses[31, 33, 35]. Sensitization with noble metal such as palladium can further enhance the sensory mechanism by influencing the rate of response, the type of the response, and the magnitude of the response[35, 62].

Depending on the sensitizer and environment, chemical sensitization, electrical sensitization, or a joint action may be present. In chemical sensitization, the noble metal promoter activates a test gas and promotes its subsequent oxidation (catalysis). In electrical sensitization, the promoter modifies the electrical behaviour of the semiconductor by a direct charge transfer interaction[35, 62]. While palladium is often found to be in nanoparticulate form[36] some fraction may also diffuse into the $TiO_2$ matrix[63].

From the optical perspective, conductometric variations due to the modulation of the free carrier concentration should show up as changes in the absorbed light intensity at near infrared wavelengths (NIR), due to Drude's theory of free carrier interaction[64-66]. Depending on the state of palladium in the $TiO_2$ matrix, different interactions may also be possible. If palladium is in the form of metal particles distributed throughout the $TiO_2$ matrix, these particles would then dissolve hydrogen to form palladium hydride. The formation of palladium hydride will affect the conductivity of the particles in addition to providing a volume expansion[36]. Diffused palladium will affect the electrical behaviour of the semiconductor by a modification of the free carrier charge density. All of these possibilities are expected to affect both the real and imaginary parts of the refractive index of the Palladium-$TiO_2$ nano-composite matrix. From the Kramers-Kronig relation, it is known that an absorptive increase should be accompanied by variations in the real part of the refractive indices[67]. Therefore, a combination of a transmission type analysis with an observation of the resonance peak of the fiber Bragg Grating is used to characterize the possible sensory mechanism, or combination thereof.

In addition, optical frequency domain reflectometry type measurements are performed to examine the potential of the developed integration method for distributed type chemical sensing. This type of characterization combines a time of flight type analysis with an analysis of the backscattered optical Rayleigh signal which may be used to characterize absorptive and refractive index variations across the length of the sensory element since a diminishing forward propagating light intensity in the fiber due to absorption by the sensory film is expected to influence the Rayleigh scattering properties. In other words, it may be used for the detection of chemical concentration variations that may exist across the sensory element since a single optical fiber sensor can function as a linear combination of hundreds of sensors.

## Experimental

**Materials and Methods**

Methods for the manufacture of $SnO_2$, $TiO_2$, and $SiO_2$ 3D nanomaterials, using non-ionic Pluronic type block-copolymers, were adapted from the literature. The manufacture of the ZnO nanostructure was obtained by exploration as the formation of 3D amorphous ZnO nanostructures with the triblock-copolymer route is not well explored, therefore, prior publications on this were not found. Precursor solutions for $TiO_2$ were prepared with the following general procedure. First, a highly concentrated Pluronic F-127 solution was prepared by mixing 10g 1-butanol with 1.3g 37%HCl and 1.3g $H_2O$, to which 5g of the bloc-polymer was added and dissolved. This solution was subsequently used as the polymer source. Various concentrations of Titanium isopropoxide in 1-butanol were prepared to which measured quantities of the polymer source were added to attain the following molar ratios of titanium isopropoxide:Pluronic F-127:37%HCl:$H_2O$:1-butanol. Ti-1 1:0.004:0.54:3.15:28.4, Ti-2 1:0.0066:0.83:4.43:29.3, Ti-3 1:0.01:1.12:5:30.4, Ti-4 1:0.013:1.43:5.64:31.58, Ti-5 1:0.016:1.43:2.3:32.47. The mixtures were stirred at room temperature for 2 hours and left to settle for one day before use. For the preparation of palladium doped $TiO_2$, 0.32g of $PdCl_2$ was dissolved in 6g of 1-butanol and 0.76g of 37% HCl. From this an amount required to obtain a 3mol% ratio to Ti was subsequently added to the prepared Ti-4 mixture. Dense $TiO_2$ was obtained without any polymer with a solution of titanium isopropoxide in 2-methoxyethanol, stabilized with 2:1 moles of ethanolamine. The manufacture of the $SnO_2$ and $SiO_2$ precursors followed a similar route with the following molar ratios of $SnCl_4$:Pluronic F-127:37%HCl:ethanol(Sn-1 1:0.008:2:21.7, Sn-2 1:0.04:7.7:39.6) and tetraethyl orthosilicate:Pluronic F-127:37%HCl:ethanol(Si-1 1:0.0084:6.8:51.2, Si-2 1:0.0122:8:54.3, Si-3 1:0.0149:10.28:63.3, Si-4 1:0.0066:4.57:49.74, Si-5 1:0058:3.99:49.74, Si-6 1:0.0051:3.43:45.22, Si-7

1:0.0045:2.86:45.22). The preparation of the precursor solution for producing a 3D ZnO nanostructure consisted of the following ratios of zinc acetate dihydrate:Pluronic F-127:30%NH$_4$OH:ethanol, 1:0.0174:18.8:33.4.

The precursor solutions were deposited on ~2x2cm square pieces of <100> silicon wafer by the spin cast method. 100μL of the precursor solutions were deposited after which the samples were quickly accelerated to 2500RPM and held for 30s. The annealing procedure consisted of heating the samples in a furnace to 400-800°C at a heating/cooling rate of 1°C/minute, and holding for 2 hours at the final temperature.

The optical properties of the samples were characterized by a Jobin Yvon spectroscopic Ellipsometer. For the SnO$_2$, SiO$_2$ and ZnO nanostructures, the Lorentz oscillator model provided good fits to the measured data[68]. For the TiO$_2$ nanostructures the New Amorphous model was used to obtain the refractive indices[69]. Variations in the refractive indices were represented by changes in the porosity, modelled using Bruggeman's effective medium theory.

In order to write fiber Bragg gratings in D-shaped fiber, hydrogen loading was performed for 2 weeks at 1600 psi. A type II fiber Bragg grating (FBG) was inscribed by a phase mask (2.5x1cm with 1060nm period) with a 248nm KrF laser source (GSI Lumonics PM-844) with a cumulative fluence of ~6,000 pulses at ~50mJcm-2[70, 71].

The preparation of the nanomaterial integrated optical fiber consisted of first the removal of a 4μm cladding material on a 15cm section of the flat side of the fiber(21minutes in 5NH4F:1HF, HF requires proper handling as it is an extremely dangerous acid), in order to expose the fiber core. The fiber was pulled through a petri dish at ~5mm/s containing the precursor solution, to coat the etched portion. Several coatings were applied and after each coating the fiber was left to dry for 30minutes. After coating, the fiber was left to dry overnight followed by its placement in a 6mm in diameter quartz tube positioned in a tube furnace. The ends of the quartz tube containing the fiber were sealed with rubber ferules with additional connections to allow for the flow of various gasses.

## Results and Discussion

### Nanomaterial Refractive Index Exploration

In **Figure 4A** the Ellipsomery determined refractive indices of TiO$_2$ prepared by precursor solutions Ti-1 to Ti-5, annealed at 400°C in air, along with refractive indices for a solution processed dense TiO$_2$, are shown. The refractive indices by CVD were obtained from the literature[56, 57]. **Figure 4B** shows the refractive indices of the prepared SiO$_2$ precursor solutions, with increasing refractive indices from low to high. The processed precursor solutions for SnO$_2$ and ZnO provided the refractive indices shown in **Figure 4C**. The polymer concentration has the effect demonstrated in **Figure 4D** on the refractive indices. The precursor solutions T-1 through Ti-4 provide decreasing refractive index values, whereas the refractive indices of samples prepared by Ti-5 were increasing. The precursor solution Ti-4 provides an approximate location of the minimum refractive index that is achievable with the current method. Although, a more detailed study is needed to confirm, it is suspected that this behaviour is linked with the central symmetry typical of phase diagrams associated with block-copolymers. The complete removal of the block-copolymer at temperatures ~350°C densify the nanomaterials with subsequent heating at higher temperatures. Therefore, an annealing study was performed to explore the effect of the annealing temperature on the refractive indices (**Figure 4C**). It is observed that samples annealed at 800°C had refractive indices that were ~7% greater. Therefore, not too significant of change was observed in the refractive indices due to densification by annealing at higher temperatures at which the block-copolymer support structure is no longer present. In **Figure 5** example prepared films annealed at 600°C are shown to demonstrate the high optical quality. In general, the scale of the nano-features are <10nm, therefore, there is a scale separation of 40λ which should result in minimal scattering for light at the visible and NIR frequencies.

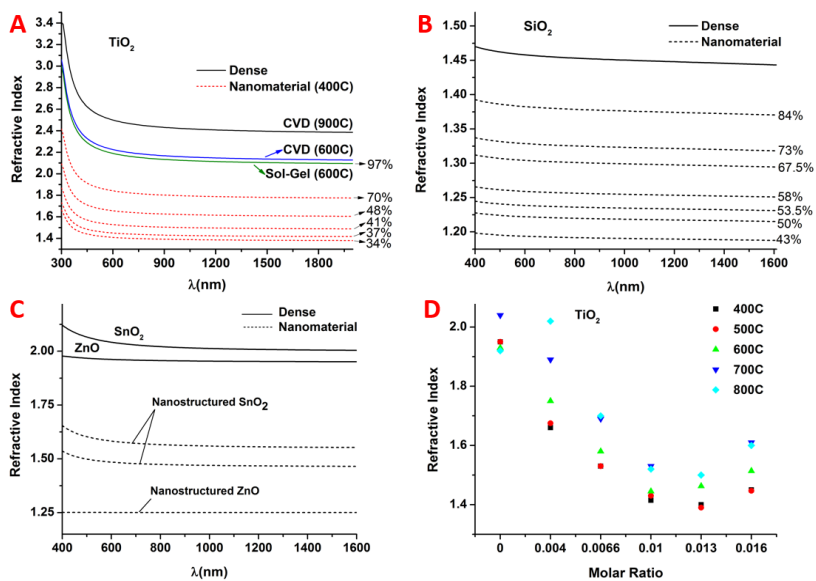

**Figure 4. A:** TiO$_2$ refractive indices provided by precursors Ti-1 through Ti-5 along with the refractive indices obtained for the dense TiO$_2$ precursor. In addition, refractive indices obtained from literature for TiO$_2$ deposited by CVD are included for comparison. **B:** SiO$_2$ refractive indices obtained from precursors Si-1 through Si-7. **C:** SnO$_2$ and ZnO refractive indices obtained by the developed precursor solutions. **D:** Examination of variations in the refractive indices as a function of annealing temperature and block-copolymer to Ti molar ratio.

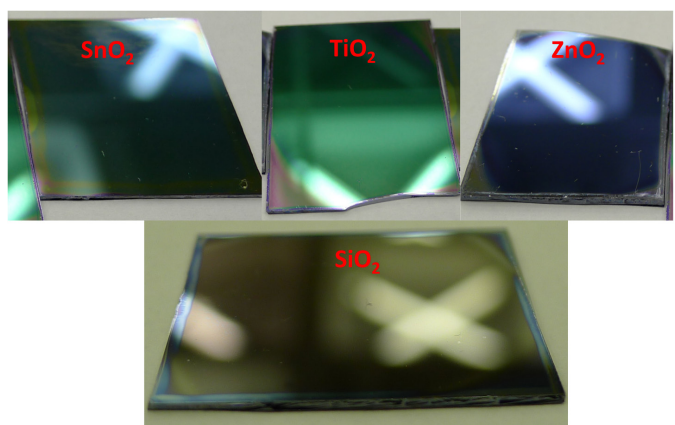

**Figure 5:** Photographs showing the examples of the prepared nanomaterial oxides on silicon wafer. The films were obtained with precursors providing the lowest refractive indices.

**Manufactured Nanomaterials**

To examine the structure of the various nanomaterials precursor solution Ti-4, Sn-2, and the solution for ZnO were deposited on TEM grids with 9x9 windows of 0.1x0.1mm, with 50nm thick silicon nitride membranes. Bright field images were obtained with a JEOL JEM2100F TEM for the various nanostructures (**Figures 6A-6C**). A large degree of porosity is evident with an average grain size of <10nm for $TiO_2$ and $SnO_2$. On the other hand, the bright field TEM for ZnO shows an average grain size ~15-20nm. The precursor formulation of ZnO is highly basic as opposed to the acidic formulations of the other oxides. With Pluronic type block-copolymers, an acidic formulation is typical. The developed base stabilized precursor for ZnO appears to provide nanostructures with shapes similar to that of the acid stabilized precursors, with the exception of an overall increase in the formed grain dimension. In **Figure 6D** an SEM image of a ZnO sample prepared on silicon wafer is shown for comparison. With TEM imaging, the grains appear as the darker regions whereas in SEM imaging, they appear as lighter regions.

**Nanomaterial Anti-Reflection Coating**

The refractive indices and film thicknesses provided by the transfer matrix method coupled with the minimization techniques to obtain global and omnidirectional reflectivity minimums, were used to manufacture two-layer $TiO_2$ and nanostructured $SiO_2$ thin-film coating on crystalline silicon wafers. Two coatings of the solution containing titanium isopropoxide, monoethanolamine, and 2-methoxyethnaol with molar ratio of 1:2:30 were spin cast at 3000RPM to obtain the dense $TiO_2$ layer. After each coating, the samples were placed on a hotplate set to 300°C, for 10 minutes, followed by the deposition of the $SiO_2$ precursor providing a refractive index in the vicinity of 1.22. The as prepared samples were then annealed at 600°C, as described before. With Ellipsometry the thickness of the dense $TiO_2$ layer in contact with the silicon substrate was measured to be 59nm and the thickness of the $SiO_2$ layer was measured to be 107nm. Practically, no measurable difference was noted in the reflectance of the two-layer coating measured at angles of incidence of 17° and 42°. From the measured reflectance spectra the average reflectivity was estimated to be ~3%. This compares reasonably well with the design value of 2.82%, given that there is some degree of uncertainty in estimating the refractive indices of porous nanomaterials by Ellipsometry due to modelling and the slight differences in the obtained film thicknesses.

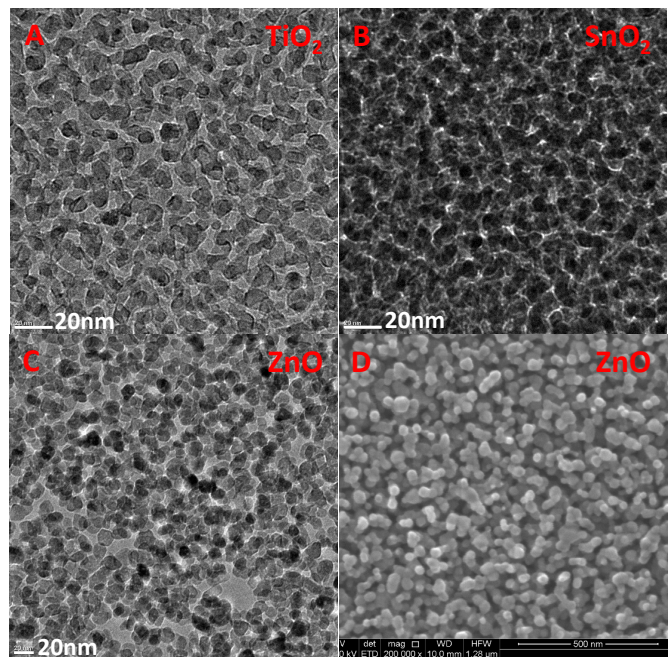

**Figure 6:** Bright field TEM images for **A:** $TiO_2$ **B:** $SnO_2$, and **C:** ZnO prepared on silicon nitride TEM grids **D:** SEM image or the surface of a ZnO sample prepared on silicon wafer. The images correspond to precursors yielding the lowest refractive indices.

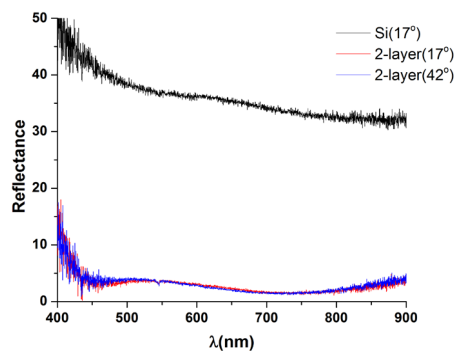

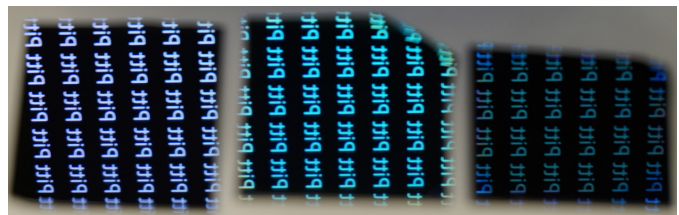

**Figure 7: Top:** Measured reflectance of crystalline silicon and a two-layer anti-reflection coating. **Bottom:** Photograph of an image projected onto crystalline silicon, a one-layer $TiO_2$ coating, and the manufactured two-layer coating, showing reductions in the reflected visible light.

**Pd-Doped TiO$_2$ Hydrogen Sensor**

The cross sectional SEM image of **Figure 8** illustrates an example of a Pd-doped TiO$_2$ nanomaterial coating on D-shaped optical fiber, with a nanomaterial refractive index of ~1.46 after annealing in air, with a film of ~2μm in thickness. The annealing procedure was performed as described before and the Pd-doped TiO$_2$ nanomaterial coated D-shaped fiber was fastened in a quartz tube at 600°C and exposed to 0.5% hydrogen balanced with nitrogen. It is believed that in air the added palladium exists as PdO nanoparticles homogenously response as a function of temperature, with a rate change at about 400°C. Exposing the sensor to different hydrogen distributed throughout the TiO$_2$ matrix. After exposure to hydrogen, these PdO nanoparticles are reduced[36] and some of the palladium may diffuse into the TiO$_2$ lattice[63]. The dynamics of the sensor were explored from 700°C down to 400°C by cycling between 0.5% hydrogen in nitrogen and pure nitrogen, as shown in **Figure 9**. Light was coupled into the D-shaped optical fiber from a broadband light source (MPB EBS-7210) with appreciable optical power from 1515nm to 1615nm. The NIR light propagated through the 15cm sensory portion of the fiber and collected on the other side with an InGaAs photo-detector. It is observed that at 700°C the fabricated sensor had a stronger and faster response in comparison with lower temperatures. The magnitude of the response was characterized from 200°C to 700°C, as shown in **Figure 10**. It is interesting that an almost linear increase is observed in the strength of the concentrations provided the data shown in **Figure 11**. It is likely that an electronic type behaviour may be primarily responsible for the observed sensory mechanism which, to some extent, is due to the disassociation of hydrogen and its subsequent dissolution into the Pd nanoparticles. This can then be associated with a modulation in the free carrier concentration subsequently altering the measured light intensity. **Figure 12** shows the recorded photo-detector voltages when heating form 25°C to 600°C at a rate of 10°C/minute, showing strong temperature dependant sensitivity. Therefore, a test was performed to examine whether the sensory response is due to thermal conductivity variations. Helium has a thermal conductivity that is comparably high to that of hydrogen and no response was noted when the sensor was exposed to helium and recovered in nitrogen. One reason behind the temperature dependant increase in the transmitted optical power could be due to an increase in the catalytic ionosorption of oxygen, trapping conduction electrons[31, 72], although other mechanisms can produce temperature dependant conductivity variations in materials.

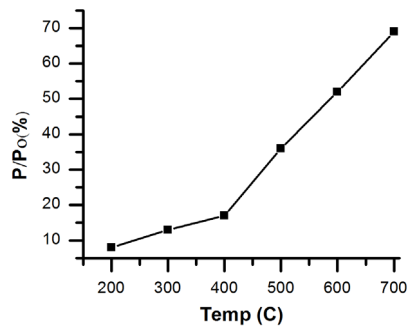

Figure 10: Response magnitude of the constructed sensor as a function of temperature when exposed to 0.5% hydrogen in nitrogen.

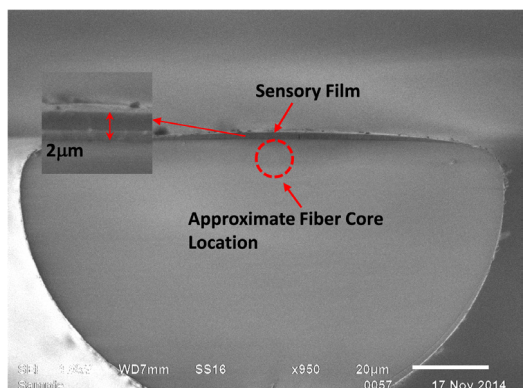

Figure 8: Cross sectional SEM of a Pd-doped TiO$_2$ nanomaterial coated D-shaped optical fiber, showing a film thickness of ~2μm in the vicinity of the fiber core.

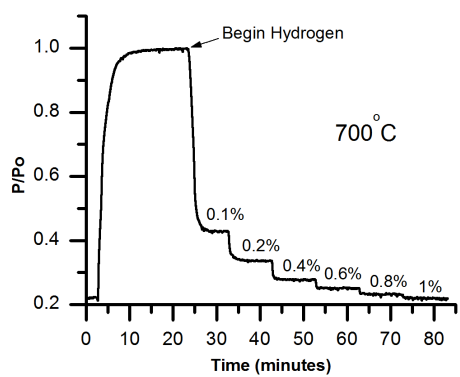

Figure 11: Normalized transmitted optical power (measured photo-detector voltage) as a function of hydrogen concentration at 700°C.

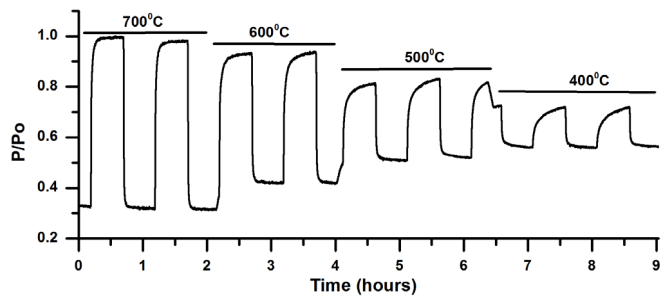

Figure 9: The dynamics of the 3mol% Pd-doped TiO$_2$ nanomaterial coated D-shaped optical fiber when exposed to 0.5% hydrogen in nitrogen and recovered with nitrogen, at temperatures 700°C to 400°C.

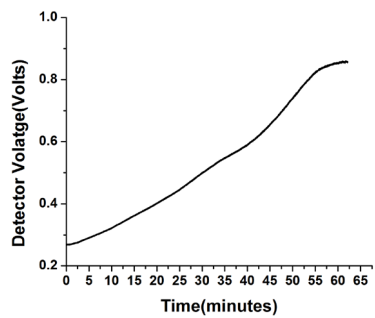

Figure 12: Photo-detector voltage during heat up from 25°C to 600°C in air at a rate of ~10°C/minute.

At the input end of the fiber a fiber optic circulator facilitated the separation of light reflected by the fiber Bragg grating from the forward propagating component. A fiber Bragg grating is very sensitive to changes in the environmental refractive indices and will have associated resonant wavelength shifts($\Delta \lambda_{peak} = 2\Lambda\Delta n_{eff}$). The in-fiber FBG facilitates the observation of changes in the real refractive index in the Pd-doped TiO$_2$ nano matrix. The formation of palladium hydride is known to cause a lattice expansion in palladium. According to the Kramers-Kronig relation, an absorptive change will always be accompanied by a change in the real part of the refractive index. In examining the reflected resonance peak of the fiber Bragg grating when exposed to 2.5% hydrogen, no observable wavelength shift in the resonance peak is detected (**Figure 13**). The dimension of the TiO$_2$ grains are observed to be <10nm with bright field TEM. It is known that the modulated space charge region is ~1-10nm, therefore, it is reasonable to assume that the TiO$_2$ matrix is fully or near fully modulated when cycled between hydrogen and nitrogen. The lack of a resonant wavelength shift indicates that there is not a measurable change in the real part of the refractive index of the nanomaterial, given the sensitivity of the instruments used. The optical spectrum analyser used in identifying the resonant wavelength has a resolvability limit of the order of ~10pm. This, combined with the confinement factor predicted by simulation gives a detection limit of ~5x10$^{-4}$ for observing refractive index variations.

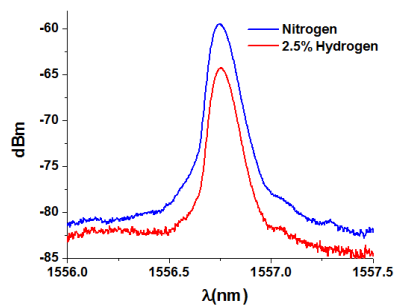

**Figure 13:** Resonance peak of the fiber Bragg grating before and after exposure to 2.5% hydrogen in nitrogen.

**Distributed Hydrogen Sensor**

A distributed type characterization was performed with Luna OBR 4600, using optical frequency domain reflectometry. **Figure 14** highlights Rayleigh backscatter measurements coupled with a time of flight type analysis for a 10cm section of the constructed sensor as a function of sensor length. The initial data was collected with a spatial resolution of 0.1mm which was subsequently reduced to 1cm, to smooth the data. Upon exposure to hydrogen, the modulated free carrier density is accompanied by changes in the absorbed NIR light. A diminishing forward propagating light intensity (Beer-Lambert law) has a direct relationship with the observed backscattered Rayleigh amplitude. The inset of **Figure 14** shows the magnitude of the backscattered Rayleigh amplitude for the highlighted section of the fiber, when exposed to various concentrations of hydrogen in nitrogen. The irregular features in the signal are believed to result from film thickness variations and possible thermal variations along the sensor length. Towards the end of the sensor, the Rayleigh signal weakens as it approaches the noise floor. The observed data provides direct evidence for the possibility of measuring concentration gradients with the developed sensor with particularly interesting applications in hydrogen fuel cells in which gradients are believed to exist due to variations in the fuel consumption across the length of the cell. This type of analysis coupled with the demonstrated nano-engineering enabled optical fiber sensor could serve as an internal monitoring and control system for fuel cells and other types of energy conversion devices.

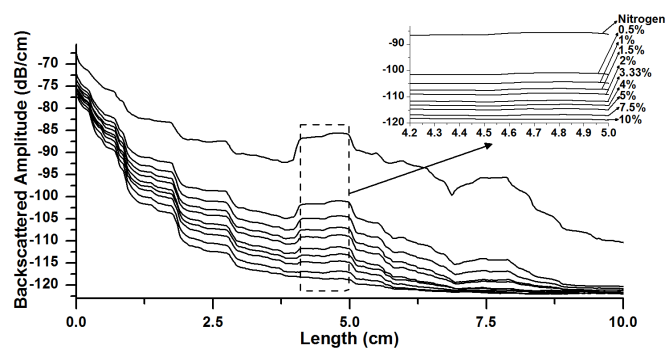

**Figure 14:** Optical frequency domain reflectometry type distributed analysis of the backscattered Rayleigh amplitude for a 10cm section of the Pd-doped TiO$_2$ nanomaterial coated D-shaped optical fiber at 600°C, exposed to various concentrations of hydrogen in nitrogen and recovered in nitrogen.

## Conclusions

A method to engineer the refractive indices of functional materials by 3D nanostructuring in the deep sub-wavelength regime is demonstrated. The method is based on block-copolymer templating coupled with a low-cost wet processing approach to provide functional optical materials with tailorable refractive indices in the range of 1.17 to 2.2 with nanostructure features <20nm. The capability to tailor the refractive indices has potential high impact applications in a variety of fields.

In the explored application of anti-reflection coating, it is shown that with a two-layer coating light reflections at a silicon air interface were reduced from ~38% to 3%. Further reductions in the reflected light are possible by surface structuring and with a combination of increasing the highest achievable refractive index of $TiO_2$ (currently 2.2) and by reducing the lowest achievable refractive index of $SiO_2$ (currently 1.17). These improvements would allow for the coating of additional layers, providing further reductions in the reflected light between silicon air interfaces. Other substrates, such as aluminium nitride, may have less stringent requirements. The developed method is easily compatible with large scale deposition methods such as ultrasonic spray coating, dip coating, etc. Techniques such as spray deposition could enable the realization of continuous refractive index gradients by altering the concentration of the block-copolymer in real time.

The tuning of the refractive indices of functional sensory materials is demonstrated to have important applications in optical fiber sensing. Optical fiber is a highly sought after sensory platform due to its small form and its capability to withstand a wide array or harsh environments. While a Pd sensitized $TiO_2$ nanomaterial served the purpose of demonstrating the applicability of the developed method, more advanced high temperature sensing materials are anticipated to offer further opportunities for sensor optimization. As demonstrated, the functional integration of sensory materials with optical fiber can provide sensory solutions for a variety of demanding applications, such as at high temperatures. The demonstrated fiber in-line technique is readily compatible with existing interrogation technologies. As shown, there is a strong potential for the detection of chemical gradients by combining existing techniques such as optical frequency domain reflectometry with the developed nano-engineering enabled optical fiber sensor. Functional metal oxide sensory materials have seen decades of exploration and a variety of methods exists to tailor their sensitivity, selectivity, response types from absorptive, refractive index, and fluorescence based. When combined with optical fiber, this merger provides a powerful combination for the detection of a variety of chemical and biological species and their gradients in a variety of harsh and sensitive environments.

As recently demonstrated, the transformation optics design methodology can be used to design various conceptual optical components, such as efficient light distribution, bending, and coupling devices for light wave/silicon photonic circuits[2]. Transformation optics is a tensor based design technique and, therefore, it generally provides designs requiring three-dimensionally distributed strong refractive index gradients. It is needless to say that such designs are very difficult, if not impossible, to manufacture with existing technologies. It may be possible to realize many transformation optics designs by combining solution based refractive index engineering with such techniques as dip pen lithography, or other compatible technologies. With such a combination, 3D refractive index gradients could be printed with a possible resolution down to 50nm.

## Acknowledgements

This work was supported by the National Science Foundation (CMMI-1054652, and CMMI-1300273) and the Department of Energy (DE-FE0003859)).

## Notes and references

[a]Department of Electrical and Computer Engineering, University of Pittsburg, Pittsburgh, PA, USA 15261.
[b]National Energy Technology Laboratory, 626 Cochrans Mill Road, Pittsburgh, PA, USA 15236.
Contact Author: zslpoole@gmail.com

1. H. F. Ma and T. J. Cui, *Nat Commun*, 2010, **1**.
2. Q. Wu, J. P. Turpin and D. H. Werner, *Light-Sci Appl*, 2012, **1**.
3. J. B. Pendry, D. Schurig and D. R. Smith, *Science*, 2006, **312**, 1780-1782.
4. D. Schurig, J. B. Pendry and D. R. Smith, *Opt. Express*, 2006, **14**, 9794-9804.
5. J. Q. Xi, M. F. Schubert, J. K. Kim, E. F. Schubert, M. F. Chen, S. Y. Lin, W. Liu and J. A. Smart, *Nat Photonics*, 2007, **1**, 176-179.
6. F. C. Y. Zhao, Q. Shen, and L. Zhang, *Progress In Electromagnetics Research*, 2014, **145**, 39-48.
7. Y. F. Huang, S. Chattopadhyay, Y. J. Jen, C. Y. Peng, T. A. Liu, Y. K. Hsu, C. L. Pan, H. C. Lo, C. H. Hsu, Y. H. Chang, C. S. Lee, K. H. Chen and L. C. Chen, *Nat Nanotechnol*, 2007, **2**, 770-774.
8. C. Lee, S. Y. Bae, S. Mobasser and H. Manohara, *Nano Lett.*, 2005, **5**, 2438-2442.
9. H. Park, D. Shin, G. Kang, S. Baek, K. Kim and W. J. Padilla, *Advanced Materials*, 2011, **23**, 5796-5800.
10. Y. Kanamori, M. Sasaki and K. Hane, *Opt. Lett.*, 1999, **24**, 1422-1424.
11. C. H. Sun, W. L. Min, N. C. Linn, P. Jiang and B. Jiang, *Appl. Phys. Lett.*, 2007, **91**.
12. S. Wang, X. Z. Yu and H. T. Fan, *Appl. Phys. Lett.*, 2007, **91**.
13. J. W. Chen, B. Wang, Y. Yang, Y. Y. Shi, G. J. Xu and P. Cui, *Appl. Opt.*, 2012, **51**, 6839-6843.
14. B. Paivanranta, T. Saastamoinen and M. Kuittinen, *Nanotechnology*, 2009, **20**.
15. S. Chhajed, M. F. Schubert, J. K. Kim and E. F. Schubert, *Appl. Phys. Lett.*, 2008, **93**.
16. M. L. Kuo, D. J. Poxson, Y. S. Kim, F. W. Mont, L. K. Kim, E. F. Schuhert and S. Y. Lin, *Opt. Lett.*, 2008, **33**, 2527-2529.
17. G. F. Fernando, D. J. Webb and P. Ferdinand, *Mrs Bull*, 2002, **27**, 359-364.
18. P. R. Ohodnicki, M. P. Buric, T. D. Brown, C. Matranga, C. J. Wang, J. Baltrus and M. Andio, *Nanoscale*, 2013, **5**, 9030-9039.
19. Z. Yibing, G. Keiser, C. Marzinsky, A. M. Schilowitz, S. Limin and A. B. Herhold, Applications of optical fiber sensors in the oil refining and petrochemical industries, 2011.

20. K. Young Ho, K. Myoung Jin, R. Byung Sup, P. Min-Su, J. Jae-Hyung and L. Byeong Ha, *IEEE Sens. J.*, 2011, **11**, 1423-1426.
21. B. Culshaw, *J. Lightwave Technol.*, 2004, **22**, 39-50.
22. C. Perrotton, R. J. Westerwaal, N. Javahiraly, M. Slaman, H. Schreuders, B. Dam and P. Meyrueis, *Opt. Express*, 2013, **21**, 382-390.
23. S. F. Silva, L. Coelho, O. Frazao, J. L. Santos and F. X. Malcata, *IEEE Sens. J.*, 2012, **12**, 93-102.
24. X. Tang, J. Provenzano, Z. Xu, J. Dong, H. Duan and H. Xiao, *J. Mater. Chem.*, 2011, **21**, 181-186.
25. X. Tang, K. Remmel, X. Lan, J. Deng, H. Xiao and J. Dong, *Anal. Chem.*, 2009, **81**, 7844-7848.
26. X. Wei, T. Wei, J. Li, X. Lan, H. Xiao and Y. S. Lin, *Sensor. Actuat. B-Chem.*, 2010, **144**, 260-266.
27. Q. Yan, S. Tao and H. Toghiani, *Talanta*, 2009, **77**, 953-961.
28. M. Yang, J. Dai, X. Li and J. Wang, *J. Appl. Phys.*, 2010, **108**, -.
29. J. Zhang, X. Tang, J. Dong, T. Wei and H. Xiao, *Opt. Express*, 2008, **16**, 8317-8323.
30. S. K. Mishra and B. D. Gupta, *Analyst*, 2013, **138**, 2640-2646.
31. M. E. Franke, T. J. Koplin and U. Simon, *Small*, 2006, **2**, 36-50.
32. G. Korotcenkov, *Mat. Sci. Eng. B-Solid.*, 2007, **139**, 1-23.
33. A. Ponzoni, E. Comini, I. Concina, M. Ferroni, M. Falasconi, E. Gobbi, V. Sberveglieri and G. Sberveglieri, *Sensors-Basel*, 2012, **12**, 17023-17045.
34. P. R. Solanki, A. Kaushik, V. V. Agrawal and B. D. Malhotra, *Npg Asia Mater*, 2011, **3**, 17-24.
35. N. Yamazoe, *Sensor. Actuat. B-Chem.*, 1991, **5**, 7-19.
36. M. K. Kumar, L. K. Tan, N. N. Gosvami and H. Gao, *J. Appl. Phys.*, 2009, **106**.
37. I.-D. Kim, A. Rothschild, B. H. Lee, D. Y. Kim, S. M. Jo and H. L. Tuller, *Nano Lett.*, 2006, **6**, 2009-2013.
38. R. Clausius, *Die mechanische U'grmetheorie*, 1879, **2**, 62.
39. O. F. Mossotti, *Mem. di mathem. e fisica in Modena.*, 1850, **24 11**, 49.
40. M. Garnett, *Philosophical Transactions of the Royal Society of London. Series A, Containing Papers of a Mathematical or Physical Character*, 1904, **203**, 385-420.
41. D. A. G. Bruggeman, *Annalen der Physik*, 1935, **416**, 636-664.
42. T. C. Choy, *Effective medium theory : principles and applications*, Clarendon Press ; Oxford University Press, Oxford England New York, 1999.
43. N. A. K. Meznarich and University of Michigan., Dissertation, 2012, p. 107 p.
44. M. C. Orilall and U. Wiesner, *Chemical Society Reviews*, 2011, **40**, 520-535.
45. M. P. Tate, V. N. Urade, S. J. Gaik, C. P. Muzzillo and H. W. Hillhouse, *Langmuir*, 2010, **26**, 4357-4367.
46. S. Shao, M. Dimitrov, N. Guan and R. Kohn, *Nanoscale*, 2010, **2**, 2054-2057.
47. V. N. Urade and H. W. Hillhouse, *J. Phys. Chem. B*, 2005, **109**, 10538-10541.
48. P. Yang, D. Zhao, D. I. Margolese, B. F. Chmelka and G. D. Stucky, *Chem. Mater.*, 1999, **11**, 2813-2826.
49. Y. M. Lam, N. Grigorieff and G. Goldbeck-Wood, *Phys Chem Chem Phys*, 1999, **1**, 3331-3334.
50. W. H. Southwell, Google Patents, 1986.
51. J. Zhao and M. A. Green, *IEEE Trans. Electron Devices*, 1991, **38**, 1925-1934.
52. M. F. Schubert, F. W. Mont, S. Chhajed, D. J. Poxson, J. K. Kim and E. F. Schubert, *Opt. Express*, 2008, **16**, 5290-5298.
53. Y. J. Chang and Y. T. Chen, *Opt. Express*, 2011, **19**, A875-A887.
54. X. Guo, H. Y. Zhou, S. Guo, X. X. Luan, W. K. Cui, Y. F. Ma and L. Shi, *Opt. Express*, 2014, **22**, A1137-A1144.
55. M. Born and E. Wolf, *Principles of optics : electromagnetic theory of propagation, interference and diffraction of light*, Cambridge University Press, Cambridge ; New York, 1999.
56. M. A. Green, *Sol Energ Mat Sol C*, 2008, **92**, 1305-1310.
57. E. D. Palik, *Handbook of optical constants of solids*, Academic Press, Orlando, 1985.
58. B. S. Richards, *Sol Energ Mat Sol C*, 2003, **79**, 369-390.
59. A. Hartung, S. Brueckner and H. Bartelt, *Opt. Express*, 2010, **18**, 3754-3761.
60. Z. L. Poole, P. Ohodnicki, R. Z. Chen, Y. K. Lin and K. P. Chen, *Opt. Express*, 2014, **22**, 2665-2674.
61. B. E. A. Saleh and M. C. Teich, *Fundamentals of photonics*, Wiley-Interscience, Hoboken, N.J., 2007.
62. J. Moon, J. A. Park, S. J. Lee, T. Zyung and I. D. Kim, *Sensor Actuat B-Chem*, 2010, **149**, 301-305.
63. G. Altamura, L. Grenet, C. Roger, F. Roux, V. Reita, R. Fillon, H. Fournier, S. Perraud and H. Mariette, *J Renew Sustain Ener*, 2014, **6**.
64. R. V. Baltz and W. Escher, *Physica Status Solidi B-Basic Research*, 1972, **51**, 499-&.
65. H. Peelaers, E. Kioupakis and C. G. Van de Walle, *Appl. Phys. Lett.*, 2012, **100**, 011914-011913.
66. P. R. Ohodnicki, M. Andio and C. Wang, *J. Appl. Phys.*, 2014, **116**.
67. J. D. Jackson, *Classical electrodynamics*, Wiley, New York, 1999.
68. H. J. Yvon, http://www.horiba.com/fileadmin/uploads/Scientific/Downloads/OpticalSchool_CN/TN/ellipsometer/Classical_Dispersion_Model.pdf.
69. H. J. Yvon, http://www.horiba.com/fileadmin/uploads/Scientific/Downloads/OpticalSchool_CN/TN/ellipsometer/New_Amorphous_Dispersion_Formula.pdf.
70. M. L. Åslund, J. Canning, M. Stevenson and K. Cook, *Opt. Lett.*, 2010, **35**, 586-588.
71. S. Bandyopadhyay, J. Canning, P. Biswas, M. Stevenson and K. Dasgupta, *Opt. Express*, 2011, **19**, 1198-1206.
72. A. K. Singh, S. B. Patil, U. T. Nakate and K. V. Gurav, *Journal of Chemistry*, 2013, **2013**, 8.